\documentclass{aastex63}

\usepackage{amsmath,amsfonts,amssymb}
\newcommand{\ud}{\,\mathrm{d}}
\newcommand*{\rom}[1]{\uppercase\expandafter{\romannumeral #1\relax}}
\received{2019 December 10}
\revised{2020 May 31}
\accepted{2020 June 5}

\submitjournal{ApJ}

\shorttitle{Thermodynamical evolution of SADs}
\shortauthors{Xue et al.}

\begin{document}

\title{Thermodynamical Evolution of Supra-Arcade Downflows}

\correspondingauthor{Hui Li}
\email{nj.lihui@pmo.ac.cn}
\correspondingauthor{Yang Su}
\email{yang.su@pmo.ac.cn}

\author[0000-0003-4829-9067]{Jianchao Xue}
\affil{Key Laboratory of Dark Matter and Space Astronomy, Purple Mountain Observatory, Chinese Academy of Sciences,\\
10 Yuanhua Road, Nanjing 210023, China}
\affil{School of Astronomy and Space Science, University of Science and Technology of China,\\
96 Jinzhai Road, Hefei 230026, China}

\author[0000-0002-4241-9921]{Yang Su}
\affiliation{Key Laboratory of Dark Matter and Space Astronomy, Purple Mountain Observatory, Chinese Academy of Sciences,\\
10 Yuanhua Road, Nanjing 210023, China}
\affiliation{School of Astronomy and Space Science, University of Science and Technology of China,\\
96 Jinzhai Road, Hefei 230026, China}

\author[0000-0003-1078-3021]{Hui Li}
\affiliation{Key Laboratory of Dark Matter and Space Astronomy, Purple Mountain Observatory, Chinese Academy of Sciences,\\
10 Yuanhua Road, Nanjing 210023, China}
\affiliation{School of Astronomy and Space Science, University of Science and Technology of China,\\
96 Jinzhai Road, Hefei 230026, China}

\author[0000-0003-2875-7366]{Xiaozhou Zhao}
\affiliation{Centre for mathematical Plasma Astrophysics, Department of Mathematics, KU Leuven,\\
Celestijnenlaan 200B, 3001 Leuven, Belgium}
 
\begin{abstract}

Supra-arcade downflows (SADs) are dark, teardrop-shaped features descending upon flare arcades. They are thought to be the results of magnetic reconnection, but the detailed formation mechanism of SADs and their relationship to flare energy release are still unclear. In this work, we explore the thermodynamical properties of SADs in the 2011 October 22 limb flare using the observations of the Solar Dynamics Observatory/Atmospheric Imaging Assembly and an improved Differential Emission Measure (DEM) technique. Different heating events around SADs are identified and the propagation speeds of plasma heating are calculated. The first heating event starts with the increase of the emission measure at temperatures higher than $5\,\mathrm{MK}$, about 2.8 minutes before the arrival of the first studied SAD. Its propagation speed is about $140\,\rm km\,s^{-1}$, a little faster than the speed of the SAD. However, the other two events have fast propagation speeds more than $700\,\rm km\,s^{-1}$. We suspect that the first heating event can be explained by adiabatic compression, but the others may have different causes. Besides, we observed that SADs can push away their surrounding spikes. The formation of SADs is further explained on the basis of patchy and bursty magnetic reconnection that reconnection outflows may push away surrounding plasma and leave dark lanes behind them. The reliability of the DEM results, heating and cooling mechanisms, and other SAD explanations are discussed.
\end{abstract}

\keywords{Solar flares (1496); Solar flare spectra (1982); Solar coronal heating (1989); Solar magnetic reconnection (1504);}

\section{Introduction} \label{sect:intro}
Supra-arcade downflows \citep[SADs;][]{1999ApJ...519L..93M} are dark extended or void features in a supra-arcade fan descending upon the post-flare arcades. They are mostly observed during the decay phase of long-duration event (LDE) flares \citep{2000SoPh..195..381M} seen in soft X-rays, high-temperature extreme ultraviolet (EUV) lines and white light \citep{1999ApJ...519L..93M, 2002ApJ...579..874S, 2003SoPh..217..247I, 2010ApJ...722..329S}, while sometimes also seen in the impulsive phase of solar flares \citep{2004ApJ...605L..77A, 2007A&A...475..333K}. It is generally agreed that SADs are voids with high-temperature plasma depletion, according to spectroscopic observations \citep{2003SoPh..217..247I} and differential emission measure (DEM) techniques \citep{2012ApJ...747L..40S, 2014ApJ...786...95H, 2017A&A...606A..84C}. Until now, no observation has indicated that SADs have higher temperature than their surrounding plasma. The SADs' thermal pressure is suspected to be lower than their surroundings, which makes it difficult to explain an SAD's formation and relatively long lifetime (a few minutes to 20 minutes).

SADs are interesting because they are generally thought to be the results of magnetic reconnection; the latter is a principal mechanism responsible for types of solar activities. The heating of post-flare loops has been well explained by chromospheric evaporation \citep{Benz2016}. However, the highest temperatures are at the base of the current sheet \citep{2018ApJ...854..122W}, or in a supra-arcade fan when facing the current sheet as we will show in this work, but the plasma heating in this region is poorly understood. Studying the relationship between plasma heating and SADs may help us understand the energy release during flares. \citet{2004ApJ...605L..77A} and \citet{2007A&A...475..333K} found that many SADs are related to hard X-ray bursts, and suspected that they are associated with energy release processes. \citet{Reeves2017} found that the temperatures of the regions with SADs tend to increase but the temperatures of the regions without SAD decrease, and the heating mechanism was interpreted as adiabatic compression in front of SADs. Recently, \citet{2019arXiv191005386R} studied plasma heating in the current sheet using a three-dimensional (3D) magnetohydrodynamic (MHD) simulation and concluded that adiabatic compression is important for plasma heating in the late phase of the eruption.

Explanations have been proposed for SAD formation on the basis of magnetic reconnection, but none of them can explain all the properties of SADs. The explanations we know are sorted as follows.
\begin{enumerate}
\item SADs as the cross-sections of evacuated flux tubes \citep{1999ApJ...519L..93M, 2006ApJ...642.1177L, 2009ApJ...697.1569M, 2010ApJ...722..329S}, or wakes behind retracting loops \citep{2012ApJ...747L..40S}. From the initial stage of their observations, SADs are treated as outflows from magnetic reconnection, and magnetic flux provides supporting pressure and prevents SADs from being filled in immediately. In this scenario, SADs and supra-arcade downflow loops (SADLs) are thought to be the same phenomena but observed from different directions \citep{2009ApJ...697.1569M, 2010ApJ...722..329S}; The former are visible when the current sheet is observed face-on and the latter are seen when observing the current sheet edge-on. However, \citet{2009ApJ...697.1569M} explained the brightening ahead of SADLs by chromospheric evaporation, but \citet{2011ApJ...742...92W} wondered that why retracting flux tubes are not filled with hot plasma by chromospheric evaporation.  When \citet{2012ApJ...747L..40S} found the brightening ahead of SADs, they reinterpreted SADs as wakes behind retracting flux tubes. \citet{2013ApJ...776...54S, 2016ApJ...831...94S} simulated a flux tube shrinking in unreconnected loops, and their results suggest that heating due to compression and cooling due to rarefaction are expected to be observed before and behind a supersonic flux tube, respectively. But \citet{2013ApJ...775L..14C} argued that the wakes behind flux tubes should be filled by surrounding plasma rapidly.
\item SADs as voided cavities formed by blast wave expansion triggered by bursty reconnection events \citep{2011A&A...527L...5M, 2015ApJ...807....6C, 2016ApJ...832...74Z}. This model can explain the low SAD densities, different SAD sizes, and interactions between SADs. But the SAD temperature is more than twice the temperature of their surroundings, contrary to the DEM results \citep{2014ApJ...786...95H} with the observations of the Solar Dynamics Observatory (SDO)/Atmospheric Imaging Assembly (AIA) and the Hinode/X-Ray Telescope.
\item SADs as outflow jets from reconnection sites penetrating into the denser flare arcades \citep{2013ApJ...775L..14C}. In this model, the density stratification of the solar corona is emphasized, and continuous reconnection is assumed to explain the long-time existence of SADs. However, intermittent reconnection outflows, including SADs, and quasi-periodic pulsations (QPPs) in solar flares are widely reported \citep{2004ApJ...605L..77A, 2013ApJ...767..168L, 2015ApJ...807...72L, 2016ApJ...830L...4S, 2018ApJ...866...64C}, which may suggest that the magnetic reconnection is bursty. 
\item SADs as the results of Rayleigh-Taylor instability (RT instability) between the reconnection outflows and denser supra-arcade fan \citep{2014ApJ...796L..29G, 2014ApJ...796...27I}. In this model, the reconnection is not necessary to be either patchy or bursty. This model can explain the splits of SAD heads when SADs interact with bright spikes \citep{2014ApJ...796...27I}. However, in the MHD simulations of RT instability \citep{2007ApJ...671.1726S, 2014ApJ...796L..29G, 2016ApJ...825L..29X}, more than one bubble, SAD-like structure, occurs at the same time and nearby locations, which is contrary to the observations that SADs are usually localized and intermittent. Furthermore, the speeds of the simulated SADs \citep[about $50\,\rm km\,s^{-1}$,][]{2014ApJ...796L..29G} are lower than the initial speeds of SADs from observations \citep[usually higher than $100\,\rm km\,s^{-1}$,][]{2011ApJ...730...98S}.
\item SADs as the outflow jets of localized and intermittent fast magnetic reconnection \citep{2009ApJ...707..420S}. Using 3D simulation, \citet{2009ApJ...707..420S} produced SAD-like magnetic structures by localized and intermittent fast magnetic reconnection. But they did not synthesize emission maps to compare with observations, and did not explain the low emissivity of SADs. 
\end{enumerate} 

In this work, we aim to explore more thermodynamical properties of SADs using a different DEM method with AIA observations and try to explain the associated heating and cooling phenomena. By combining SAD observations and the SAD models of predecessors, we may further understand SADs. The observations and data reduction are introduced in Section~\ref{sect:meth}. In Section~\ref{sect:res}, we present our observation results and analyses. The reliability of DEM results, the mechanisms of plasma heating and cooling, and SAD formation mechanisms are discussed in Section~\ref{sect:dis}. A summary of this work is given in Section~\ref{sect:con}.
%%%%%%%%%%%%%%%%%%%%%%%%%%% Method %%%%%%%%%%%%%%%%%%%%%%%%%%%
\section{Observations and Data Reduction} \label{sect:meth}
AIA \citep{2012SoPh..275...17L} is one of the three payloads of the SDO \citep{2012SoPh..275....3P}. In its seven EUV channels, AIA takes narrowband images with a temporal resolution of $12\,\rm s$ and pixel size of $0.6''$. In this work, we will analyze SADs in an M1.3 Geostationary Operational Environmental Satellite (GOES) class flare that occurred on 2011 October 22. This SAD event was widely studied due to the relatively clear supra-arcade fan and large SADs \citep{2012ApJ...747L..40S, 2014ApJ...786...95H, 2014ApJ...796...27I, Reeves2017}. The SADs are clearly seen in the AIA $131\,\rm \AA$ wave band, which is mainly contributed by Fe~\rom{21} and corresponds to plasma temperature of about $10\,\rm MK$.

 The SDO/AIA data are processed using the standard procedure \emph{aia\_prep.pro} in the SolarSoft Ware (SSW) package. The modified version of the sparse inversion code \citep{2015ApJ...807..143C,2018ApJ...856L..17S} is adopted to extract plasma DEMs from the AIA EUV channels except for the $304 \,\rm \AA$. This code can effectively constrain the emission measure (EM) distribution at high temperatures above a few MK, and in the mean time, keep the reconstructed data consistent with the observed data. Before the DEM calculation, $2\times2$ pixels are binned to improve the signal-to-noise ratio. The AIA temperature response functions are obtained by \emph{aia\_get\_response.pro}, with EVE normalization (keyword \emph{evenorm}) and $94 \,\rm \AA$ channel correction (keyword \emph{chiantifix}) enabled. The EM, in units of $\mathrm{cm}^{-5}$, is defined by
\begin{equation}\label{eq:em}
\mathrm{EM}=\int n_\mathrm{H} n_e \ud l \approx n_e^2 L \,,
\end{equation}
where $n_\mathrm{H}$ and $n_e$ are the number densities of hydrogen and electron, and $l$ and $L$ are the path lengths along the line of sight (LOS). Since the DEM inversions are carried out with logarithmic temperature spacing (the interval is $\Delta \log T/\rm{K} = 0.05$), the EM-weighted temperature $T_{\mathrm{EM}}$ is also derived in logarithmic scale \citep{2015ApJ...807..143C}:
\begin{equation} \label{eq:avet}
\log T_\mathrm{EM} = \frac{\sum_{i=a}^b \mathrm{EM}_i \log T_i}{\sum_{i=a}^b \mathrm{EM}_i} \, ,
\end{equation}
where $\mathrm{EM}_i$ is the emission measure of $\log T_i$ bin, $a$ and $b$ are lower and upper sum limits, respectively. Analogically, the $\mathrm{DEM_i}$ is calculated by 
\begin{equation} \label{eq:dem}
\mathrm{DEM}_i = \frac{\mathrm{EM}_i}{\Delta T_i} = \frac{\mathrm{EM}_i}{\frac{\mathrm{d}T_i}{\mathrm{d}\log T_i}\,\Delta\log T} = 
\frac{\mathrm{EM}_i}{T_i \ln 10\,\Delta\log T} \, .
\end{equation}

%%%%%%%%%%%%%%%%%%%%%%%%%%% Results %%%%%%%%%%%%%%%%%%%%%%%%%%%
\section{Results}  \label{sect:res}

%%%%%%%%%%% 2011-10-22 %%%%%%%%%%%
\begin{figure}
\plotone{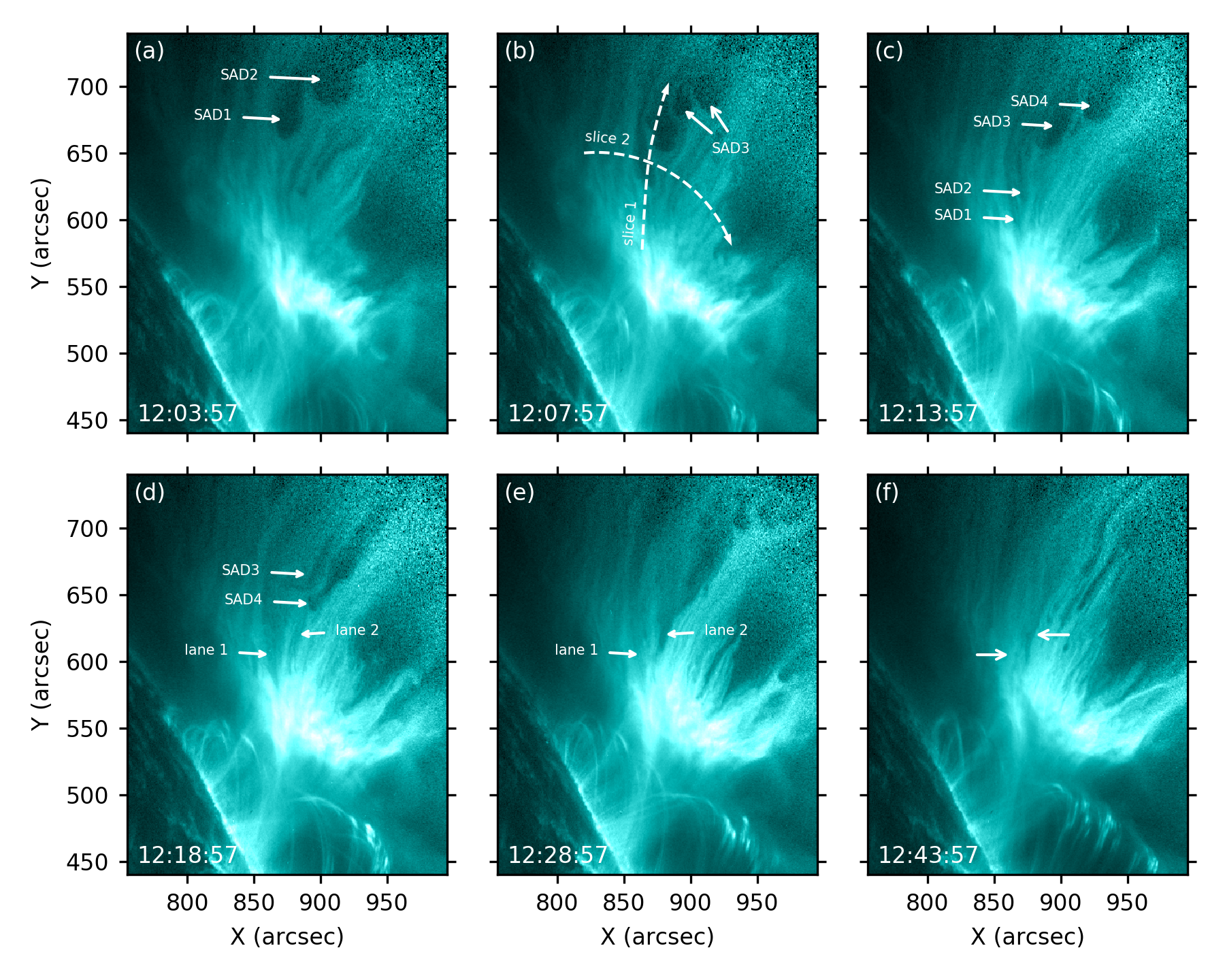}
\caption{\label{fig:1022overview}
AIA $131 \,\rm \AA$ maps. The four marked SADs and two lanes are traced. The two arrows in (f) point to the same positions as those in (e). The two slices in (b) show the positions of time--distance plots in Figures~\ref{fig:1022slice1} and \ref{fig:1022slice2}. The off-disk emission is enhanced for clarity.}
\end{figure}

\begin{figure}
\plotone{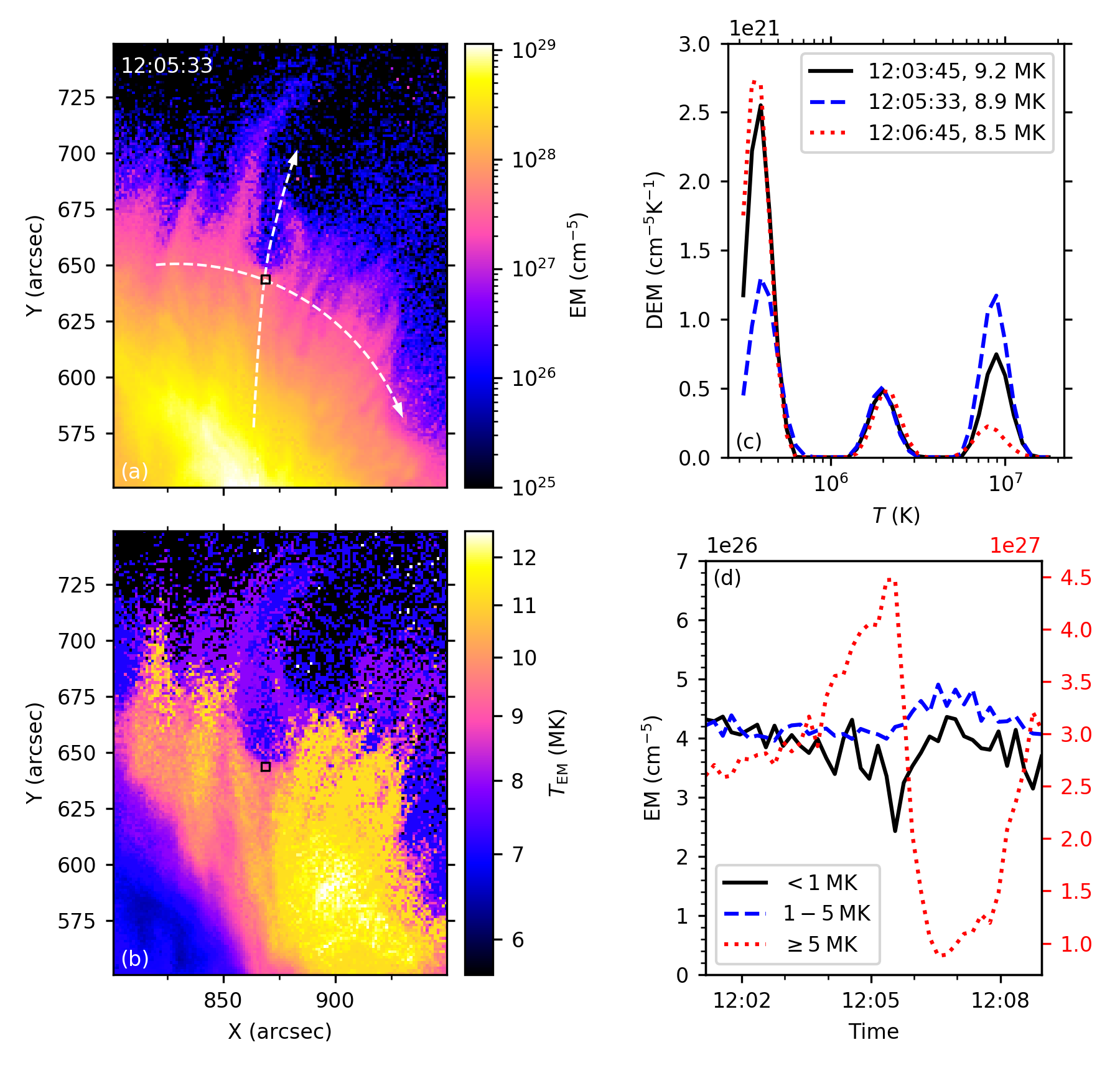}
\caption{\label{fig:1022dem}
(a)-(b) EM (a) and $T_{\mathrm{EM}}$ (b) maps by integrating $T\geq 5 \,\rm MK$ components. The square symbol marks a $3.6\times3.6 \,\mathrm{arcsec}^2$ region for detailed DEM analyses. The slices in (a) are the same as those in Figure~\ref{fig:1022overview}(b). (c) Distributions of averaged DEM along the temperature of the square region at three different times. Their observation times in AIA $131\,\rm\AA$ channel and $T_\mathrm{EM}$ of $T\geq 5 \,\rm MK$ components are noted. (d) Evolution of averaged EM in different temperature ranges of the selected region; the left $Y$-axis is for the solid black and dashed blue curves ($T<5 \,\rm MK$) and the right $Y$-axis is for the dotted red curve ($T\ge 5\,\rm MK$).}
\end{figure}

\begin{figure}
\plotone{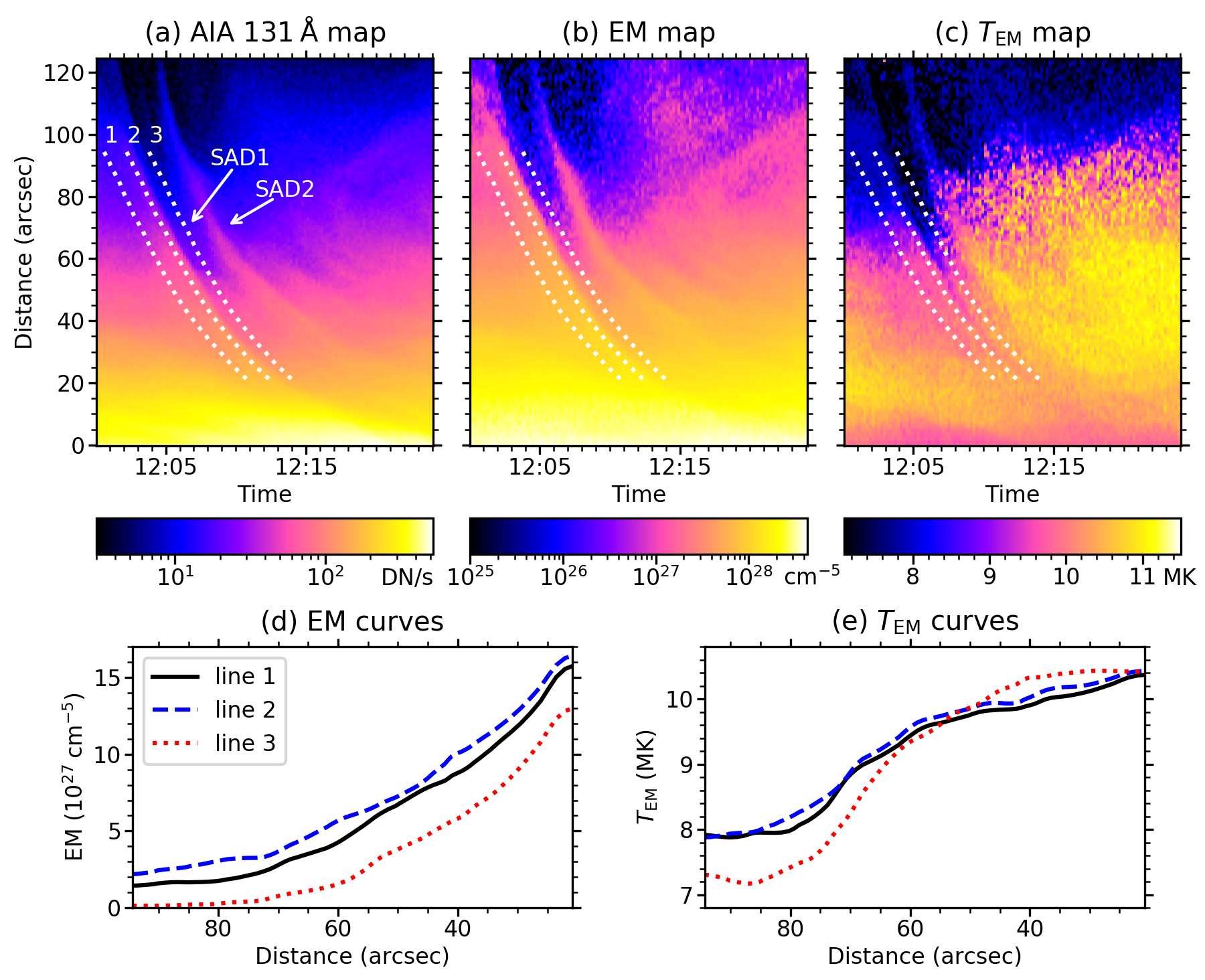}
\caption{\label{fig:1022slice1}
(a)-(c) Time--distance plots of (a) AIA $131 \,\rm \AA$, (b) EM and (c) $T_{\mathrm{EM}}$ along slice 1 in Figure~\ref{fig:1022overview}(b). SAD1 and SAD2 are marked in (a). Dotted lines 1-3 are for detailed analyses in (d)-(e). (d)-(e) Smoothed EM and $T_{\mathrm{EM}}$ distributions of lines 1-3. Only $T\geq 5 \,\rm MK$ components are considered.}
\end{figure}

\begin{figure}
\plotone{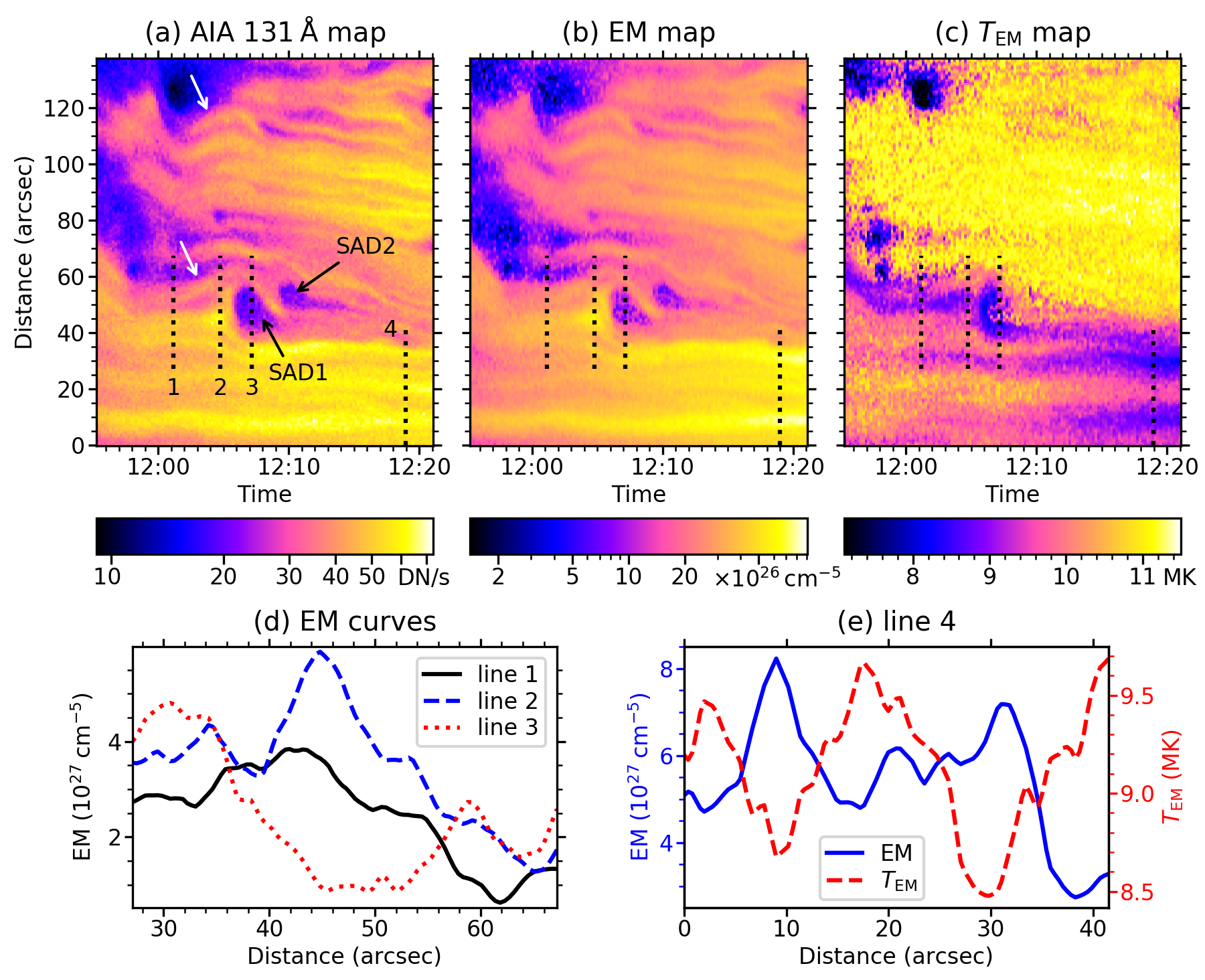}
\caption{\label{fig:1022slice2}
(a)-(c) Time--distance plots of (a) AIA $131 \,\rm \AA$, (b) EM and (c) $T_{\mathrm{EM}}$ along slice 2 in Figure~\ref{fig:1022overview}(b). Two SADs and two disturbed spikes are marked in (a) by black and white arrows, respectively. Vertical dotted lines 1-4 are for detailed analyses in (d)-(e). (d) EM distributions along lines 1-3. (e) EM (solid blue) and $T_{\mathrm{EM}}$ (dashed red) distributions along line 4. These values are averaged over three time frames. Only $T\geq 5 \,\rm MK$ components are considered.}
\end{figure}

\begin{figure}
\begin{center}
\plotone{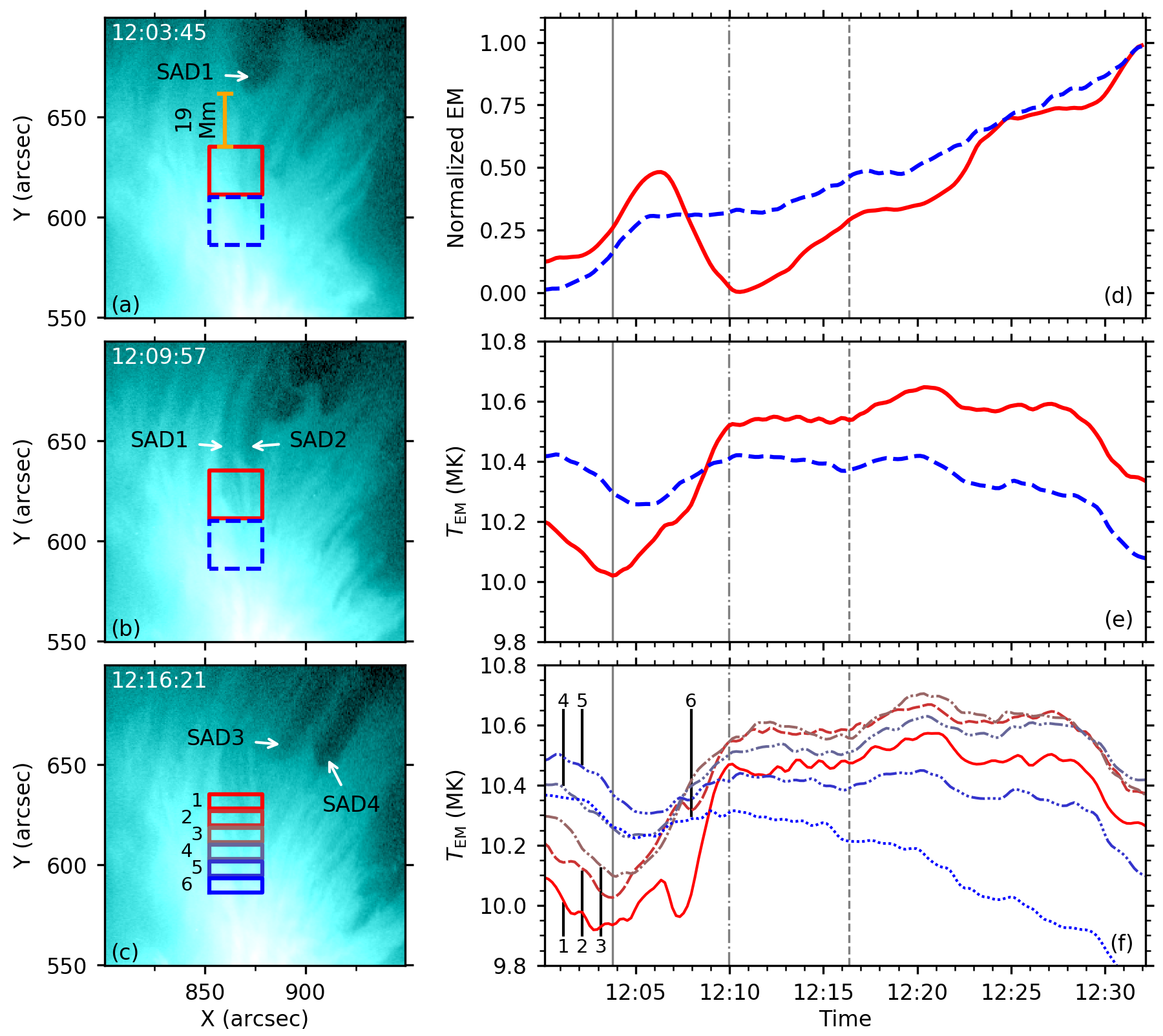}
\caption{\label{fig:1022temp}
(a)-(c): AIA $131 \,\rm \AA$ maps. The rectangle regions are for average $T_\mathrm{EM}$ and $\mathrm{EM}$ calculations; the two regions in (a)-(b) are divided into six regions in (c). The concerned SADs are marked. (d) Averaged and normalized $\mathrm{EM}$ curves of the regions marked in (a)-(b) with the corresponding colors and line styles. (e) Average  $T_\mathrm{EM}$ curves of the same region as (d). (f) Average $T_\mathrm{EM}$ curves of the regions marked in (c) with the corresponding colors and numbers. The vertical lines in (d)-(f) correspond to the observation times of images in (a)-(c). Curves in (d)-(f) are smoothed. Only $T\geq 5 \,\rm MK$ components are considered.}
\end{center}
\end{figure}

Four large SADs and two dark lanes are traced in the 2011 October 22 flare. Figure~\ref{fig:1022overview} shows their evolution in AIA $131 \,\rm \AA$ images with the off-disk emission enhanced. Among of them, SAD1 and SAD2 push away surrounding plasma and drop along lane 1. At the same time, they get thinner and merge in the narrow end of lane 1 at last. Lane 1 remains visible for at least $15\,\rm minutes$ (Figures~\ref{fig:1022overview}(c)-(e)), until it disappears for being squeezed especially by SAD3 and SAD4. SAD3 and SAD4 occur after SAD1 and SAD2 and drop along lane 2. Interestingly, the SAD3 head is split by a bright spike (Figure~\ref{fig:1022overview}(b)), the same phenomenon as that reported by \citet{2014ApJ...796...27I}. Meanwhile, SAD3 is slowed down and then descends following SAD4. Lane 2 gets much wider when SAD4 and SAD3 penetrate it, and remains visible for at least $22\,\rm minutes$ (Figures~\ref{fig:1022overview}(d)-(f)).

During their dropping, the front and side boundaries of the SADs are found to be brighter than the surrounding plasma. For a clear view on the brightening and physical parameter changes in SAD boundaries, we used time--distance slice and DEM inversion techniques. An example of EM and $T_{\mathrm{EM}}$ maps is shown in Figure~\ref{fig:1022dem}. The square symbols in Figures~\ref{fig:1022dem}(a)-(b) mark the intersection of slices of time--distance plots in Figures~\ref{fig:1022slice1}-\ref{fig:1022slice2}. The DEM distributions of the square region at different times are plotted in Figure~\ref{fig:1022dem}(c). The three curves in Figure~\ref{fig:1022dem}(c) correspond to the times when SAD1 starts to disturb the square region (as background), the bright front boundary, and SAD1 itself, sequentially. The DEM plots show that the plasma could be divided into three segments by temperature: cool plasma with $T<1\,\rm MK$, coronal background and foreground with $1\leq T<5\,\rm MK$, and hot plasma with $T\geq 5\,\rm MK$. Figure~\ref{fig:1022dem}(d) shows the EM evolution of the three components in the square region. The hot components increase in the front boundary of SAD1 and decrease when SAD1 crosses the area. The variations of $T<5\,\rm MK$ components are possibly unreliable, which will be discussed in Section~\ref{subs:dem}. Since the variation of the hot components is on order of $1.5\times 10^{27}\,\rm cm^{-5}$, much higher than that of $T<5\,\rm MK$ components on order of $1.5\times 10^{26}\,\rm cm^{-5}$, we suspect that the following analyses are reliable if only hot components are considered \citep{2015SoPh..290.2211G, 2016ApJ...819...56S}. With only $T\geq 5\,\rm MK$ components included, the average $T_\mathrm{EM}$ of the selected area at three different times are about 9.2, 8.9 and $8.5\,\rm MK$, as labeled in Figure~\ref{fig:1022dem}(c).

The time--distance plots in Figure~\ref{fig:1022slice1} are from the position of slice 1 marked in Figure~\ref{fig:1022overview}(b), which follows the dropping trajectory of SAD1 and SAD2. The $Y$-axis ``distance'' in Figures~\ref{fig:1022slice1}(a)-(c) starts at the base of the slice and increases along the arrow direction in Figure~\ref{fig:1022overview}(b). Three lines marked in Figures~\ref{fig:1022slice1}(a)-(c) have the same distance-coordinate and permanent time differences. Line 1 is treated as background, line 2 traces the front boundary of SAD1, and line 3 traces SAD1 itself. The EM and $T_\mathrm{EM}$ distributions along the three lines are plotted in Figures~\ref{fig:1022slice1}(d)-(e). 

From Figure~\ref{fig:1022slice1}(a), brightening in locations ahead of SAD1 and between SAD1 and SAD2 are visible. Figures~\ref{fig:1022slice1}(d)-(e) show that the SAD1 front boundary has higher EM and similar $T_\mathrm{EM}$ compared with line 1, indicating that the brightening in the front boundary of SAD1 is mainly contributed by the higher EM. SAD1 has both lower EM and lower $T_\mathrm{EM}$, consistent with the results of \citet{2014ApJ...786...95H}. However, low EM of SADs means that SADs are more sensitive to the unknown blended plasma along LOS, so it is not clear whether SADs are cooler than their surrounding plasma \citep{2017A&A...606A..84C}.  For the region between SAD1 and SAD2, Figures~\ref{fig:1022slice1}(a)-(c) show that both EM and temperature increase significantly and contribute to the brightening. 

Figure~\ref{fig:1022slice2} has a similar layout to Figure~\ref{fig:1022slice1} but is along the slice 2 in Figure~\ref{fig:1022overview}(b), and thus the side motions can be traced by bright fan spikes, as marked by the white arrows in Figure~\ref{fig:1022slice2}(a). When SAD1 is approaching, the AIA $131\,\rm \AA$ intensity and EM of the plasma ahead of SAD1 increase, see Figures~\ref{fig:1022slice2}(a)-(b) and (d), and the spikes beside the SAD1's trajectory are pushed away. This kind of motion starts at more than 4 minutes before SAD1's arrival (the spike pointed by the lower white arrow), and affects the spikes that are about $60''$ (about $43 \,\rm Mm$) away from SAD1 (pointed by the upper white arrow). However, only a few spikes in the left side of SAD1, lower side in Figures~\ref{fig:1022slice2}(a)-(c), are observed to be pushed away; but the enhancements of their AIA $131\,\rm \AA$ intensity and EM are obvious. The temperature increases are not significant until the approach of SAD2, see Figure~\ref{fig:1022slice2}(c).

In general, the AIA $131\,\rm \AA$ intensity and EM of the whole region in Figures~\ref{fig:1022slice2}(a)-(b) increase with time. However, the left region beside SAD1, i.e., the bottom part in Figure~\ref{fig:1022slice2}(c), cools down continuously. At the late time around 12:19~UT, the EM and temperature distributions along line 4 are shown in panel (e). We find that the place with higher EM tends to have lower temperature, which means that the thermal pressure along line 4 tends to achieve an equilibrium.

The temperature and EM evolution are further analyzed and shown in Figure~\ref{fig:1022temp}. Two regions on the trajectory of SAD1 and SAD2, lower than the square area in Figures~\ref{fig:1022dem}(a)-(b), are selected manually in Figures~\ref{fig:1022temp}(a)-(b). They are divided into six regions in Figure~\ref{fig:1022temp}(c). Their EM and $T_\mathrm{EM}$ curves are plotted in Figures~\ref{fig:1022temp}(d)-(f) with the corresponding colors, and the times of the left maps are shown with the vertical solid/dashed lines. The temperature of the region marked by solid red lines in panels (a)-(b) starts increasing at 12:03:45~UT when the SAD1 is about $19\pm0.9 \,\rm Mm$ away and $168\pm24\,\rm s$ before its arrival, as shown in panels (a) and (e). The temperature increase gets faster at about 12:07~UT and almost stops when SAD2 reaches the edge of the red region (Figure~\ref{fig:1022temp}(b)). At around 12:16:21~UT when SAD3 and SAD4 are approaching (Figure~\ref{fig:1022temp}(c)), the temperature starts to increase again. The temperature of the dashed blue region in Figures~\ref{fig:1022temp}(a)-(b) starts increasing at about 12:06~UT, $132\pm24\,\rm s$ later than the red region. If their heating is due to a same cause, which propagates along the SAD lane, considering that the distance of the upper edges of the two regions is about $18\,\rm Mm$, the propagation speed of plasma heating is $137\pm20\,\rm km\,s^{-1}$. This speed is a little faster than the SAD1's dropping speed of $113\pm17\,\mathrm{km\,s^{-1}}$. However, at times around 12:16~UT and 12:24~UT, the temperatures of both the red and blue regions increase at almost the same time. If the causes of their heating are still the same those propagating from the upper site, the heating propagation speed should be more than $700\,\rm km\,s^{-1}$, if the time interval of them being heated is less than twice the observation cadence ($24\,\mathrm{s}$). The EMs of the two regions increase quickly around 12:03~UT but do not change much around 12:16~UT and 12:24~UT (Figure~\ref{fig:1022temp}(d)), indicating that the first heating event is possibly related to the density enhancement, but the latter two are not. 

The net cooling rate of the red region is about $0.98\,\mathrm{kK\,s^{-1}}$ between 12:00:45~UT and 12:02:57~UT, and that of the blue region is $0.76\,\rm kK\,s^{-1}$ between 12:01:33~UT and 12:04:09~UT. The net heating rate of the red region is about $1.34\,\rm kK\,s^{-1}$ between 12:03:45~UT and 12:09:57~UT, and that of the blue region is $0.54\,\rm kK\,s^{-1}$ between 12:05:21~UT and 12:09:57~UT. If the cooling rates do not change during the short period, the total heating rates of the two regions at the corresponding periods should be about 2.32 and $1.30\,\rm kK\,s^{-1}$, respectively.  The detailed temperature changes can be seen in Figure~\ref{fig:1022temp}(f) for smaller regions in Figure~\ref{fig:1022temp}(c). We find that the temperature increase in a lower region is less than that of a upper region. The highest temperature slightly rises from the region 5 to regions 2-4 during the studied period. We will discuss on the heating and cooling mechanisms in Section~\ref{subs:com} and \ref{subs:cool}, respectively.

%%%%%%%%%%% Discussion %%%%%%%%%%%%%%
\section{Discussion} \label{sect:dis}
%%%%%%%%%%%%%%%%%%%%%%%
\subsection{Reliability analyses of the DEM results} \label{subs:dem}

\begin{figure}
\plotone{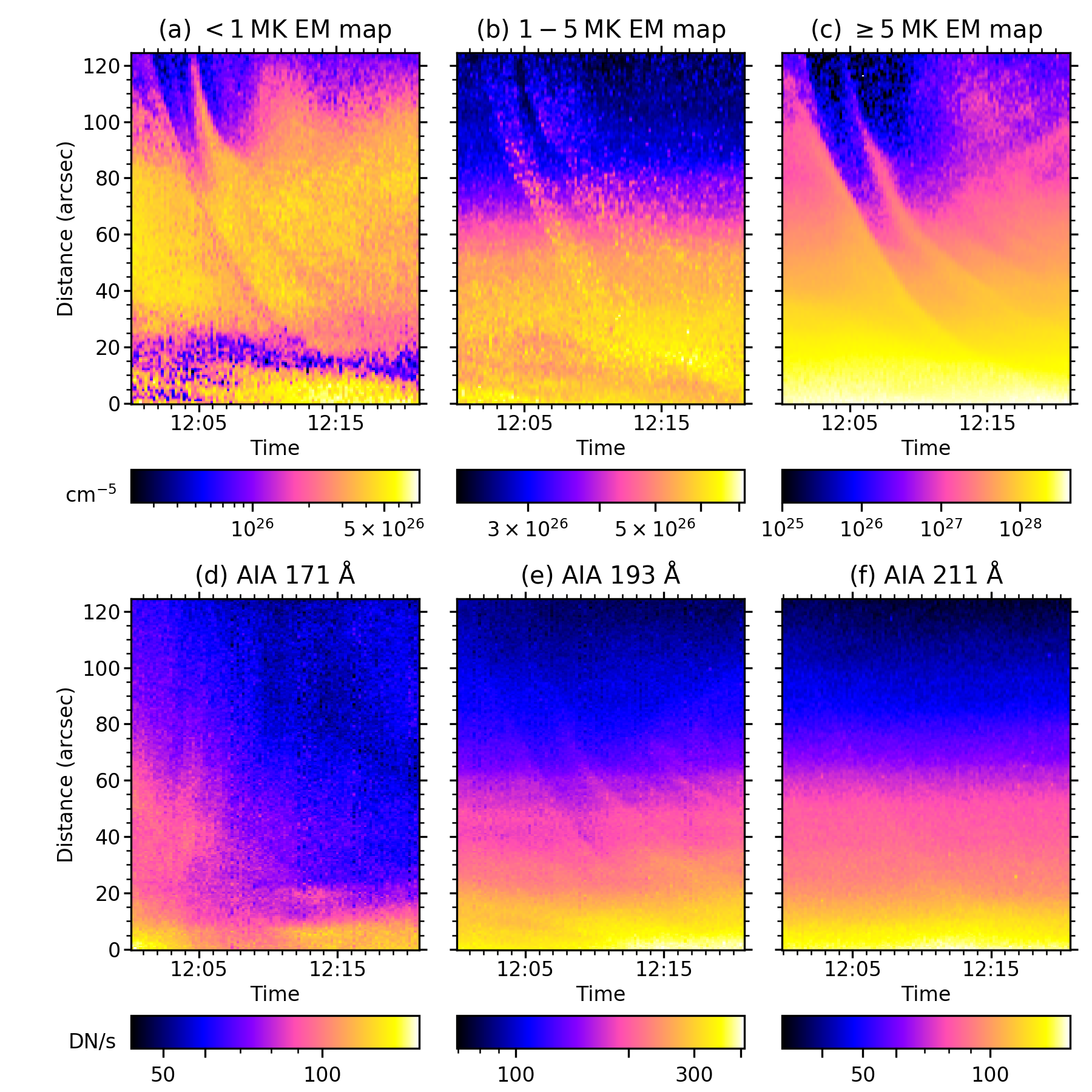}
\caption{\label{fig:check_dem}
Time--distance maps along the slice 1 in Figure~\ref{fig:1022overview}(b). (a)-(c) Summed EM maps of different temperature components. (d)-(f) AIA 171, 193, and $211\,\mathrm{\AA}$ maps sequentially.
}
\end{figure}
One of the advantages of the Sparse DEM code \citep{2015ApJ...807..143C} we used here is that it can constrain the DEM solutions so that the difference between the observed data and the reconstructed data are well limited within the given tolerance, which is relatively small in our case \citep{2018ApJ...856L..17S}. However, it does not mean that the solutions are always the correct ones, especially when the uncertainties of the data are large. Besides, DEM inversion is intrinsically a multi-solution problem. In our work, we need to use six AIA EUV channels to derive EM distribution in more than 40 temperature bins, and the solution may not be unique. Figure~\ref{fig:check_dem} shows the time--distance plots of different EM components (a)-(c) and AIA channels (d)-(f) along the slice 1 in Figure~\ref{fig:1022overview}(b). In Figure~\ref{fig:check_dem}(c), the areas with lower hot components due to SADs have larger EM in $1-5\,\rm MK$ components (panel (b)), and EM in the front boundaries of SAD1 and SAD2 increases for $T\geq 5\,\rm MK$ components but decreases for $T<1\,\rm MK$. If the DEM results are reliable, they indicate that SADs are filled with $1-5\,\rm MK$ plasma,  and the plasma with $T<1\,\rm MK$ in SAD front boundaries is heated to $\sim 10\,\rm MK$. The emission of SADs from plasma below $5\,\rm MK$ has not been reported, either by spectroscopic observations \citep{2003SoPh..217..247I} or by DEM inversions with different codes \citep{2012ApJ...747L..40S, 2014ApJ...786...95H, 2017A&A...606A..84C}. The AIA $171\,\rm \AA$ (Figure~\ref{fig:check_dem}(d)) and AIA $211\,\rm \AA$ (Figure~\ref{fig:check_dem}(f)) images, corresponding to the plasma temperature of about $0.8$ and $2\,\rm MK$, respectively, have no obvious traces of SADs. This suggests that the EM changes in $T<5\,\rm MK$ in the locations of SADs may be introduced by the DEM technique itself. We suspect that the decrease of the derived cool EM components in SAD front boundaries is due to the fact that the AIA $131\,\rm \AA$ channel is partly contributed by Fe~\rom{8}, and this result may cause the overestimate of the hot EM components. Analogously, the possibly overestimated $1-5\,\rm MK$ components of SADs in Figure~\ref{fig:check_dem}(b) may indicate that the derived hot EM components of SADs are underestimated. 

However, the enhancement of AIA $193\,\rm \AA$ intensity in the SAD front boundaries (Figure~\ref{fig:check_dem}(e)), which is due to the fact that the AIA $193\,\rm \AA$ wave band is partly contributed by Fe XXIV emission (about $18\,\rm MK$), is consistent with the enhancement of AIA $131\,\rm \AA$ as mentioned in Section~\ref{sect:res}. Because the hot EM map is consistent with AIA $131\,\rm \AA$ observations, and the summed EM components for $T\geq 5\,\rm MK$ are much higher than that $T<5\,\rm MK$ for the supra-arcade fan, we argue that the majority of our results derived from the hot components are reasonable and have tolerable errors.

In addition, it should be pointed out that the increase of EM-weighted temperature does not always mean heating of local plasma. For example, if more hot plasma with the same temperature and density moves into the examined location but is spatially separated with the preexisting plasma along LOS, then $T_{\mathrm{EM}}$ would be higher due to larger EM at this temperature (see Eqs.~(\ref{eq:em}) and (\ref{eq:avet})). The necessary consideration of flows on the heating rate calculation will be further discussed in the next section.

%%%%%%%%%%%%%%%%%%%%%%%
\subsection{Interpretations of observed heating} \label{subs:com}
\begin{figure}
\begin{center}
\plotone{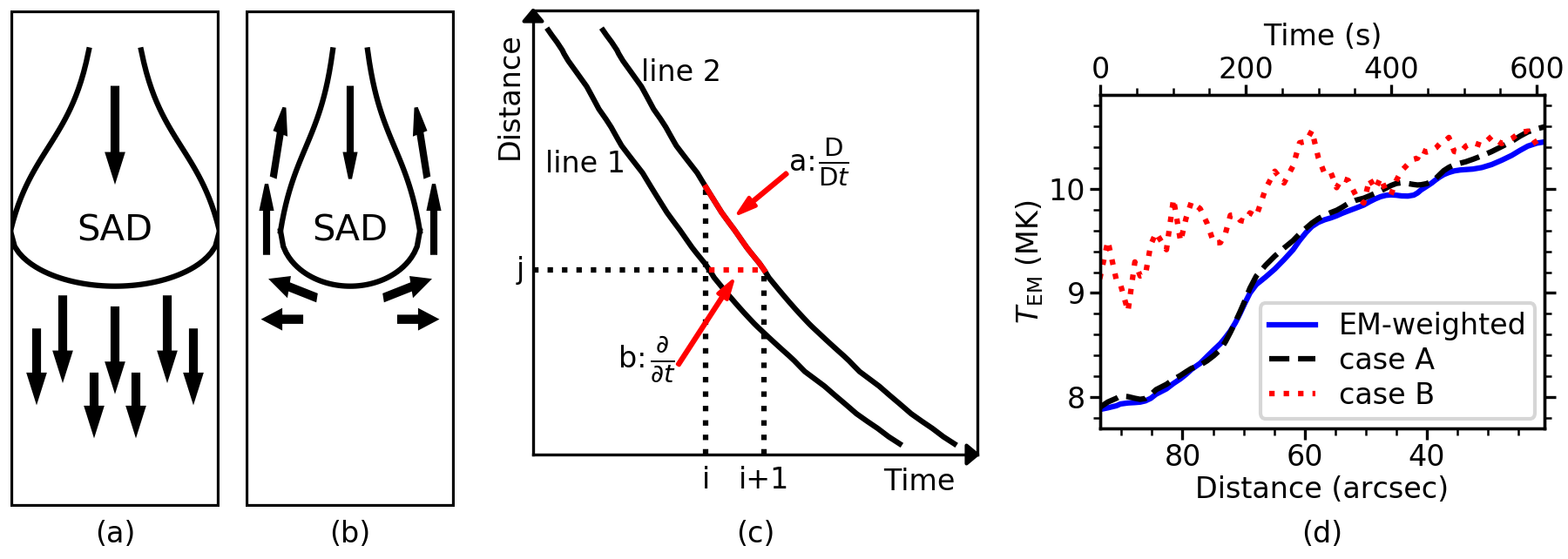}
\caption{\label{fig:1022compress}
(a)-(b) Two extreme cases. (a) Case A in which all the plasma in front of SAD is pushed down. (b) Case B in which the plasma is compressed and pushed away when the SAD is passing. Arrows mean velocities. (c) The methods of temperature calculation at time $i+1$ and position $j$ in the two cases. (d) Calculation results. The blue curve is the $T_{\mathrm{EM}}$ along line 2 as shown in Figure~\ref{fig:1022slice1}(e). The dashed black curve is obtained with Eq.~(\ref{eq:dt3}) for case A and the dotted red curve is obtained with Eq.~(\ref{eq:dt4}) for case B.}
\end{center}
\end{figure}

\citet{Reeves2017}, for the first time, calculated adiabatic heating rates associated with SADs on the basis of plane-of-sky (POS) velocities, and drew the conclusion that the heating due to adiabatic compression can overcome the conductive cooling. They derived an empirical cooling rate by DEM analyses of $0.9\,\rm kK\,s^{-1}$, the same order as our results. However, the estimated adiabatic heating rate and conductive cooling rate are on the order of $100\,\rm kK\,s^{-1}$, much higher than those derived from our DEM results. In this section, we first recalculate the adiabatic heating rate using the adiabatic equation of state and EM variations, then we analyze the possible heating mechanisms to explain the observed temperature changes.

In adiabatic condition for ideal gas, the density and temperature have the relationship of
\begin{equation}\label{eq:adia}
\frac{\mathrm{D}}{\mathrm{D}t}\left(\frac{T}{\rho^{\gamma-1}}\right) = 0 \,,
\end{equation}
where $D/Dt$ is the substantial derivative, $\rho$ is the mass density, and $\gamma=5/3$ is the polytropic index. Assuming that the path length $L$ is a constant during a short time and in a small area, considering that $\rho$ is approximately proportional to $n_e$, the variation of $T$ can be estimated by $\mathrm{EM}$ changes according to Eqs.~(\ref{eq:em}) and (\ref{eq:adia}):
\begin{equation}\label{eq:dt}
\frac{\mathrm{D}T}{T} = (\gamma-1)\frac{\mathrm{D}\rho}{\rho} \approx 
(\gamma-1)\frac{\mathrm{D}\sqrt\mathrm{EM}}{\sqrt\mathrm{EM}} \,.
\end{equation}

Physically, the change of plasma temperature is caused by (1) temperature variation at a fixed place and (2) plasma flows:
\begin{equation} \label{eq:subs}
\frac{\mathrm{D}T}{\mathrm{D}t} = \frac{\partial T}{\partial t} + (\boldsymbol{V}\nabla )T \,,
\end{equation}
where $\partial /\partial t$ is the local derivative and $\boldsymbol{V}\nabla$ is the convective derivative. Although the velocity $\boldsymbol{V}$ can be obtained by the local correlation tracking (LCT) method, small-scale flows cannot be truly identified. To evaluate the temperature changes of plasma along line 2 in Figure~\ref{fig:1022slice1}, two extreme cases are considered, as shown in Figures~\ref{fig:1022compress} (a)-(b), and the temperature calculations in the two cases are explained in Figure~\ref{fig:1022compress}(c). In case A (Figure~\ref{fig:1022compress}(a)), all the plasma in front of SAD1 drops along line 2 and does not leak, then a substantial derivative is used and Eq.~(\ref{eq:dt}) is expressed as
\begin{equation}\label{eq:dt3}
T_2^{i+1} = T_2^i + (\gamma-1)T_2^i\frac{\sqrt{\mathrm{EM}_2^{i+1}}-\sqrt{\mathrm{EM}_2^i}}{\sqrt{\mathrm{EM}_2^i}} \,,
\end{equation}
where the subscript 2 means the values of line 2, and the superscript $i$ or $i+1$ denotes time. In case B (Figure~\ref{fig:1022compress}(b)), the plasma in front of SAD1 is first compressed and then pushed away by SAD1, so only the compression of local plasma is considered, and the substantial derivative in Eq.~(\ref{eq:dt}) should be replaced by the local derivative. Then, the temperature of plasma along line 2 is estimated by
\begin{equation}\label{eq:dt4}
T_{2j} = T_{1j} + (\gamma-1)T_{1j}\frac{\sqrt{\mathrm{EM}_{2j}}-\sqrt{\mathrm{EM}_{1j}}}{\sqrt{\mathrm{EM}_{1j}}} \,,
\end{equation}
where the subscript $j$ denotes the position.

The calculations of Eqs.~(\ref{eq:dt3}) and (\ref{eq:dt4}) are performed on the smoothed EM and $T_{\mathrm{EM}}$ of line 2 and line 1; the results are shown in Figure~\ref{fig:1022compress}(d) with the dashed black and dotted red curves, respectively. The solid blue curve is $T_\mathrm{EM}$ along line 2 derived from the DEM results as shown in Figure~\ref{fig:1022slice1}(e). For case A, the calculated temperature curve (dashed black) is a little higher than the $T_{\mathrm{EM}}$ curve of line 2. For case B, local plasma is heated to a much higher temperature (dotted red). The large difference between the results of the two cases indicates that flows are critical in estimating adiabatic heating. The actual temperature variation by adiabatic compression, without including cooling terms, should be between the black and red curves, i.e., some of compressed plasma drops ahead of SAD1 and some is from the local plasma. The upper limit of the heating rate is estimated to be $\sim 10\,\rm kK\,s^{-1}$, from the maximum temperature difference between the $T_{\mathrm{EM}}$ curve of line 1 in Figure~\ref{fig:1022slice1}(e) and the red curve in Figure~\ref{fig:1022compress}(d), and the time interval between line 1 and line 2. The estimated value is about one order of magnitude higher than the net heating rate of $1.3\,\rm kK\,s^{-1}$ derived from our DEM results (Section~\ref{sect:res}).

An advantage of our calculation is that the temperature changes are derived from EM variations directly. However, the result in case A is much constricted by $T_2$, which makes the calculation questionable. We suspect that the adiabatic heating rate and conductive cooling rate calculated by \citet{Reeves2017} are probably overestimated. The authors estimated the adiabatic heating rate with POS velocities. For a region in front of an SAD, the inflow speed is approximately the speed of the SAD, but the outflow speed tends to be underestimated due to lack of characteristics. Thus the compression and adiabatic heating are overestimated. To estimate the conductive cooling rate, the authors ignored the temperatures of loop footpoints, which may result in the overestimate of the cooling rate.

We find that the propagation speed of plasma heating in front of SAD1, about $140\,\rm km\,s^{-1}$, is a little higher than the SAD1's dropping speed ($\sim 110\,\mathrm{km\,s^{-1}}$). This is new proof that supports the assumption that SADs may affect plasma heating by compression.  Our analysis is consistent with the results of \citet{2013ApJ...776...54S}, who simulated a retracting flux tube and found preheating before it. As for other heating mechanisms, MHD waves cannot explain the EM enhancements; shocks and energetic particles propagate faster than that. The propagation speed of thermal conduction is expected to be the ion-sound speed \citep[p. 679]{2005psci.book.....A} $c_s\approx \sqrt{2\gamma k_\mathrm{B} T/m_p} \approx 520\,\mathrm{km\,s^{-1}}$ for $T=10\,\mathrm{MK}$, where $k_\mathrm{B}$ is Boltzmann constant and $m_p$ is proton mass; the thermal conduction also propagates much faster than the observed plasma heating. Other proof that supports the compression between SADs and their surroundings include the following: (1) SADs have bright boundaries that are mainly contributed by the enhancement of EM (Figures~\ref{fig:1022slice1}-\ref{fig:1022slice2}), and (2) SADs are slowed down during their dropping (Figure~\ref{fig:1022slice1}), which is possibly due to drag force \citep{2006ApJ...642.1177L,2013ApJ...776...54S}.

During some heating events in the supra-arcade fan, at 12:16~UT and 12:24~UT in Figure~\ref{fig:1022temp}, the temperature increases of regions at different heights almost start simultaneously with slight EM changes. Considering the observation cadence, these heating events have the propagation speeds of over $700\,\rm km\,s^{-1}$. The fast speeds and slight EM changes cannot be explained by the SAD-related adiabatic compression. More information is necessary to determine the heating mechanism in these events; however, it is beyond the scope of this work. Our calculations are based on the assumption that the plasma heating propagates from the reconnection site and along SAD lanes, but this assumption is possibly not realistic especially after the supra-arcade fan being disturbed by SADs. However, different plasma heating behaviors (different propagation speeds, with or without obvious EM changes) suggest that different heating mechanisms may play roles on the hot supra-arcade fan.

%%%%%%%%%%%%%%%%%%%%%%%%%%
\subsection{Interpretations of observed cooling} \label{subs:cool}

\begin{figure}
\begin{center}
\includegraphics[width=3in]{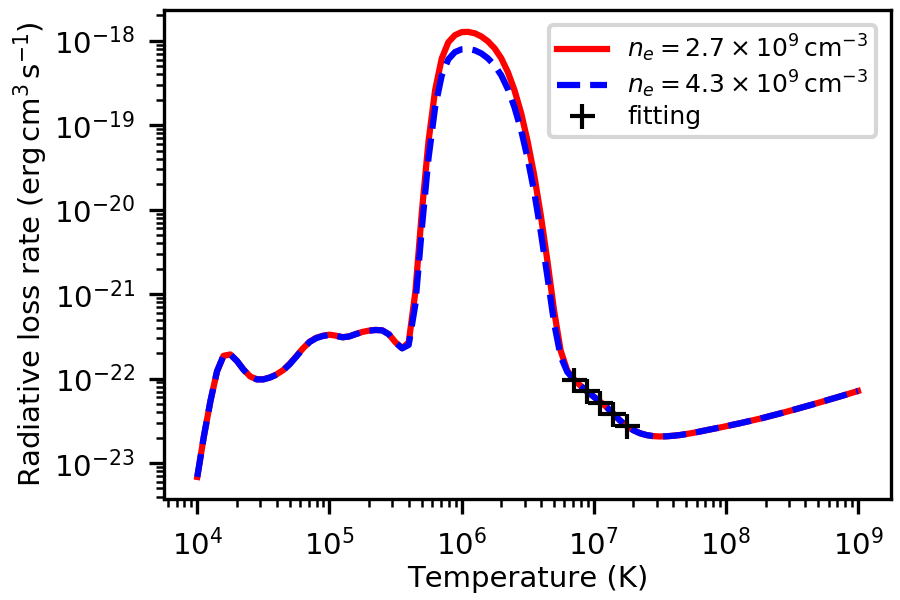}
\caption{\label{fig:rad_loss}
Radiative loss rate vs. temperature. The solid red and dashed blue curves are calculated using the coronal abundance by \citet{2012ApJ...755...33S} and the electron densities as labeled in the legend. The plus symbols show the fitting result in the temperature range of $7.08-17.8\,\mathrm{MK}$.}
\end{center}
\end{figure}

It is difficult to determine the cooling mechanisms in the supra-arcade fan after the arrival of SADs, since both cooling and heating exist simultaneously. Thus we only consider the cooling process of the solid red region in Figures~\ref{fig:1022temp}(a)-(b) between 12:00:45~UT and 12:02:57~UT, and that of the dashed blue region between 12:01:33~UT and 12:04:09~UT, with the thermal conductive cooling and radiative cooling considered. Besides, the direct calculation of the conductive cooling rate is complicate, because (1) it is usually assumed that energy propagates from the loop top to the footpoints under the effect of thermal conduction \citep{Reeves2017}, but in the regions we study here, the temperature does not decline monotonically toward the solar disk; and (2) thermal conduction is restricted by the magnetic field, but loops cannot be resolved. However, these two factors have little effect on the radiative cooling. Thus we first calculate the radiative cooling rate, then estimate the conductive cooling by making up for the derived cooling rate in Section~\ref{sect:res}. For simplicity, we ignore the potential heating.

Assuming the thickness of the supra-arcade fan to be $L=10^9\,\mathrm{cm}$ \citep{Reeves2017}, the average electron densities of the two regions during the corresponding periods are estimated to be $<n_e>_{red}=2.7\times 10^9\,\mathrm{cm^{-3}}$ and $<n_e>_{blue}=4.3\times 10^9\,\mathrm{cm^{-3}}$ (Eq.~(\ref{eq:em})), respectively. The radiative loss rate $\Lambda(T)$ is calculated using \emph{rad\_loss.pro} in the Chianti package \citep[version 8,][]{1997A&AS..125..149D,2015A&A...582A..56D} with the coronal abundance by \citet{2012ApJ...755...33S} and the electron densities of the two regions. The results are plotted in Figure~\ref{fig:rad_loss}. In the temperature range of $7.08-17.8\,\mathrm{MK}$, the two radiative loss rate functions are almost the same and can be fitted by
\begin{equation} \label{eq:rad_loss}
\Lambda(T)=\Lambda_0 T^{-1.37}=2.36\times 10^{-13} T^{-1.37} \qquad 7.08\times 10^6\,\mathrm{K} < T < 1.78\times 10^7\,\mathrm{K} \, ,
\end{equation}
as marked by the plus symbols in Figure~\ref{fig:rad_loss}. The temperatures at the regions and periods we study are fortunately within that range. With only radiative loss included, the internal energy loss is expressed as \citep[p.700]{2005psci.book.....A}
\begin{equation} \label{eq:energy_loss}
\frac{\mathrm{d}}{\mathrm{d}t}[3n_e(t)k_\mathrm{B} T(t)]=-n_e(t)^2\Lambda(T(t))\approx -n_e(t)^2\Lambda_0T(t)^{-1.37} \,,
\end{equation}
where $t$ is time. Eq.~(\ref{eq:energy_loss}) is solved to be
\begin{equation} \label{eq:T_rad}
T(t) = T_1\left(1-\frac{t-t_1}{\tau_\mathrm{rad}}\right)^{0.422} \,,
\end{equation}
where $t_1$ and $T_1$ are initial time and temperature, respectively, and the ``radiative cooling time'' $\tau_\mathrm{rad}$ is expressed as
\begin{equation} \label{eq:tau_rad}
\tau_\mathrm{rad} = \frac{1.27 k_\mathrm{B} T_1^{2.37}}{<n_e>'\Lambda_0} \approx \frac{1.27 k_\mathrm{B} T_1^{2.37}}{<n_e>\Lambda_0} \,.
\end{equation}
Plugging $T_1$ and $<n_e>$ values into Eq.~(\ref{eq:tau_rad}) yields that $\tau_{rad\_red}=3.1\,\mathrm{hours}$ and $\tau_{rad\_blue}=2.0\,\mathrm{hours}$. Using Eq.~(\ref{eq:T_rad}), the radiative cooling rates of the two regions are about $0.38\,\mathrm{kK\,s^{-1}}$ and $0.60\,\mathrm{kK\,s^{-1}}$, accounting for about 39\% of the observed cooling rate of the red region, and 79\% that of the blue one, respectively. To estimate the errors, we assume that the electron density ranges from $0.5<n_e>$ to $2<n_e>$. The corresponding radiative cooling rate of the red region is in the range of $0.19-0.77\,\rm kK\,s^{-1}$, and that of the blue region is in $0.30-1.21\,\rm kK\,s^{-1}$.

Although we cannot calculate the conductive cooling rate precisely, we could still compare the ``conductive time scales'' of the two regions approximately. The conductive time scale is defined by \citep[p.700]{2005psci.book.....A}
\begin{equation} \label{eq:tau_cond}
\tau_\mathrm{cond}=\frac{21}{5}\frac{<n_e>k_\mathrm{B} D^2}{\kappa T_1^{5/2}} \, ,
\end{equation}
where $D$ is the distance between the studied region and the loop footpoints, and $\kappa$ is the thermal Spitzer conductivity coefficient. Ignoring the difference of $D$, the ratio of the $\tau_\mathrm{cond}$ of the two regions is
\begin{equation} \label{eq:tauc_ratio}
\frac{\tau_\mathrm{cond\_red}}{\tau_\mathrm{cond\_blue}}=\frac{<n_e>_\mathrm{red}T_\mathrm{1\_blue}^{5/2}}{<n_e>_\mathrm{blue}T_\mathrm{1\_red}^{5/2}}\approx 0.66<1 \,.
\end{equation}
This means that with similar initial temperatures, the red region cools down faster than the blue region under the effect of thermal conduction. However, the effect of the radiative cooling is contrary to the conductive cooling ($\tau_{rad\_red}>\tau_{rad\_blue}$). To explain why the $T_{\mathrm{EM}}$ of the red region decreases faster than that of the blue one before SADs (Figure~\ref{fig:1022temp}(e)), the red region should be dominated by thermal conductive cooling. In addition, the differences of both the radiative and conductive cooling time scales of the two regions mainly depend on their densities, rather than temperatures, because the density variations with height are more significant than the temperature.

%%%%%%%%%%%%%%%%%%%
\subsection{Explanation of SADs} \label{subs:mod}

Explanations for SAD formation and their advantages and disadvantages have been introduced in Section~\ref{sect:intro}. Based on the observation that SADs can push away their surrounding spikes (Figures~\ref{fig:1022overview} and \ref{fig:1022slice2}), considering that the brightening around SADs is possibly caused by compression \citep{2013ApJ...776...54S, Reeves2017}, we may further interpret SADs. SADs, as the outflows of patchy and bursty magnetic reconnection \citep{2006ApJ...642.1177L, 2009ApJ...697.1569M, 2010ApJ...722..329S}, are low-density voids caused by the density stratification of the solar corona \citep{2013ApJ...775L..14C, 2014ApJ...796L..29G}; they penetrate into the denser supra-arcade fan, push away surrounding plasma, and leave dark lanes behind them. We suspect that the relatively long lifetime of SADs is mainly due to the effect of the magnetic field. On the one hand, magnetic field can inhibit plasma mixing due to the frozen-in effect \citep{2007ApJ...671.1726S}; on the other hand, magnetic tension force tends to keep field lines straight and may stop surrounding plasma from filling SADs immediately.

The splits of SAD heads (Figure~\ref{fig:1022overview}(b)) could be caused by RT instability \citep{2014ApJ...796...27I} when the magnetic fields inside an SAD are nearly parallel to the fields below it. Otherwise, RT instability is suppressed  when the magnetic fields are nonparallel \citep{2007ApJ...671.1726S}. If SADs are produced by RT instability \citep{2014ApJ...796L..29G}, more than one SAD should occur at the same time and in nearby positions, but SADs are usually observed being localized and intermittent. In the scenario of patchy reconnection, we suspect that RT instability is not necessary to produce SADs.

%%%%%%%%%%% Conclusion %%%%%%%%%%%%%%

\section{Conclusion}  \label{sect:con}

We studied the thermodynamical properties of SADs observed in the 2011 October 22 flare and focused on their bright boundaries and the implications on understanding the formation of SADs. Based on the detailed DEM analyses, we find that plasma heating starts about 2.8 minutes before the arrival of SAD1 and the heating propagation speed is around $140\,\rm km\,s^{-1}$, a little faster than the dropping speed of SAD1. The net cooling rate before the heating is $0.7-1.0\,\rm kK\,s^{-1}$ and the heating rate is $0.5-1.4\,\rm kK\,s^{-1}$. The propagation speed of plasma heating is a new evidence supporting that SADs may play a role in the heating of the supra-arcade fan through compressing ambient plasma. We estimate the plasma heating rate by the adiabatic equation of state and EM variations, and find that it should be lower than $10\,\rm kK\,s^{-1}$. However, heating phenomena with fast propagation speeds are also identified in the supra-arcade fan, which cannot be explained by adiabatic compression. We calculate the radiative cooling rates and obtain the values in the same order as that derived from the $T_{\mathrm{EM}}$ curves. We find that an upper region in the supra-arcade fan tends to be dominated by the conductive cooling due to its lower density, and a lower region tends to be dominated by the radiative cooling. By comparing DEM results and AIA observations, we find that it is better to use the hot EM components ($T\geq 5\,\rm MK$) to analyze SADs.

Considering the observation that SADs can push away their surrounding spikes, and on the basis of patchy and bursty magnetic reconnection, we further explain the formation of SADs by stating that the reconnection outflows may push away surrounding plasma and leave dark lanes behind them. The magnetic field may play an important role in inhibiting plasma mixing and preventing SADs from being filled by ambient plasma immediately. The splits of SAD heads are possibly caused by RT instability, but we suspect that RT instability is not required to produce SADs in the scenario of patchy reconnection.

Although this SAD event has been widely studied, the high supra-arcade fan and large SADs are worth a review. However, more SAD events should be analyzed to confirm the relationship between SADs and plasma heating. The relationship between SADs and SADLs should be identified to gain a further understanding of SAD formation.

\acknowledgments
AIA is an instrument on board SDO, a mission for NASA's Living with a Star program. This work is supported by NSFC grants (11427803, U1731241, U1631242, 11820101002) and by CAS Strategic Pioneer Program on Space Science, grant Nos. XDA15052200, XDA15320103, XDA15320300, and  XDA15320301. Y.S. acknowledges the Thousand Young Talents Plan, and the Jiangsu Double Innovation Plan. X.Z. is supported by a joint FWO-NSFC grant G0E9619N.

\software{DEM sparse inversion code \citep{2015ApJ...807..143C, 2018ApJ...856L..17S},
       Chianti \citep{1997A&AS..125..149D,2015A&A...582A..56D},
          Sunpy \citep{2015CS&D....8a4009S},
          Matplotlib \citep{4160265},
          Scipy \citep{scipy}}

\bibliography{SAD}{}
\bibliographystyle{aasjournal}

\end{document}